\documentclass[twocolumn]{aastex7}

\hypersetup{linkcolor=red,citecolor=blue,filecolor=cyan,urlcolor=blue}

\usepackage{graphicx,color}
\usepackage{amssymb}
\usepackage{amsmath}
\usepackage{url}
\usepackage{natbib}
\usepackage{txfonts}
\usepackage{multirow}
\usepackage{array}
\usepackage{rotating}
\usepackage{sidecap}
\usepackage{hyperref}
\usepackage{epstopdf}
\usepackage{footnote}
\usepackage{tabularx}
\usepackage{booktabs}
\usepackage{longtable}
\usepackage{tabu}
\usepackage{longtable}

\newcommand{\FeX}{\ion{Fe}{10}}
\newcommand{\FeIX}{\ion{Fe}{9}}

\newcommand{\kms}{km~s$^{-1}$}
\newcommand{\degree}{\ensuremath{^\circ}}

\newcommand{\hri}{HRI$_{EUV}$}

\begin{document}

  \title{Extreme Ultraviolet Microflashes at  Plume Bases: A Candidate for Powering the Corona and Solar Wind?}

\author[0000-0001-7620-362X]{Navdeep K. Panesar}
\affiliation{Lockheed Martin Solar and Astrophysics Laboratory, 3251 Hanover Street, Bldg. 203, Palo Alto, CA 94306, USA}
\affiliation{SETI Institute, 339 Bernardo Ave, Mountain View, CA 94043, USA}
\email{panesar@lmsal.com}
\author[0000-0001-7817-2978]{Sanjiv K. Tiwari}
\affil{Lockheed Martin Solar and Astrophysics Laboratory, 3251 Hanover Street, Bldg. 203, Palo Alto, CA 94306, USA}
\affil{Bay Area Environmental Research Institute, NASA Research Park, Moffett Field, CA 94035, USA}
\email{tiwari@lmsal.com}
\author[0000-0002-9672-3873]{Meng Jin}
\affil{Lockheed Martin Solar and Astrophysics Laboratory, 3251 Hanover Street, Bldg. 203, Palo Alto, CA 94306, USA}
\email{jinmeng@lmsal.com}

\author[0000-0002-4401-2295]{Ayla Weitz}
\affil{University of Colorado Boulder, 2000 Colorado Ave, Boulder, CO 80309, USA}
\affil{National Solar Observatory, 3665 Discovery Dr, Boulder, CO 80303, USA}
\email{Ayla.Weitz@colorado.edu}

\author[0000-0002-5691-6152]{Ronald L. Moore}
\affil{Center for Space Plasma and Aeronomic Research (CSPAR), UAH, Huntsville, AL 35805, USA}
\affil{NASA Marshall Space Flight Center, Huntsville, AL 35812, USA}
\email{ronald.l.moore@nasa.gov}
\author[0000-0002-0824-3109]{V. Aparna}
\affil{Department of Astronomy, New Mexico State University, Las Cruces, NM 88003, USA. }
\email{aparna@baeri.org}
\author[0000-0003-1281-897X]{Alphonse C. Sterling}
\affil{NASA Marshall Space Flight Center, Huntsville, AL 35812, USA}
\email{alphonse.sterling@nasa.gov}
%

\begin{abstract}

 Solar plumes – outflows of bright coronal plasma – are a major component of the open-magnetic-field corona and solar wind, but their driving mechanism remains uncertain. Here, we report on “network microflashes,” fine-scale bright bursts captured by Solar Orbiter’s Extreme Ultraviolet Imager in 174 \AA\ images encompassing magnetic network at the base of plumes. Because they sit in evidently unipolar magnetic flux, they are evidently a new, previously unidentified, kind of network event. Approximately 20 microflashes are ongoing within a plume base, with a new microflash starting every second. The energy for an average microflash is $\sim$10$^{24}$ erg, in the range of nanoflares. A 3D data-driven global MHD model yields open magnetic field with fast solar wind for the investigated plumes. From our findings, we suggest that network microflashes result from fine-scale bursts of reconnection of crossed legs of unipolar magnetic field, that the bursts are often triggered by 5-minute p-mode oscillations, and that the bursts are candidates for powering the open-field corona and solar wind. That is, unipolar microflashes such as ours are plausibly from unipolar-network-field reconnection bursts that sustain the heliosphere. 
\end{abstract}



\section{Introduction} 
The solar wind \citep{parker58} is a gush of charged particles from the Sun. It interacts with Earth’s magnetic field (magnetosphere), driving dynamic changes and influencing the planet’s atmosphere and space environment \citep{neugebauer1962}.   The solar wind  mostly originates from open magnetic field regions on the Sun and inflates the heliosphere. The fast solar wind, of speeds 450 to 1000 \kms, primarily emanates from open magnetic field regions known as coronal holes (\citealt{richardson18}, and references therein). On the other hand, much of the slow solar wind, of speeds 300--450 \kms, has been argued to originate from the edges of coronal holes that often appear next to active regions \citep{damicis15,wang19,viall20}. In either case, coronal holes are a major source of solar wind. 

Coronal plumes, in and around coronal holes, are known to be another major source of solar wind \citep{mcIntosh10,tian11,pucci14,zangrilli20}. They are primarily observed in extreme ultraviolet (EUV) emission (most clearly visible in Fe IX/X emission) and appear as bright hazy structures in on-disk coronal holes \citep{wang08,wang-y-m2016}. They stem from strong unipolar  network clumps \citep{newkirk68,avallone18}, in which smaller weak opposite minority-polarity flux  patches may also be present \citep{wang94,raouafi14,wang-y-m2016,panesar18b,avallone18,wang-y-m2022}. 
\cite{raouafi23} proposed that small-scale magnetic reconnection-driven jetting (jetlets) activity at sites of mixed-polarity flux is a major driver of coronal heating and the solar wind. A similar mechanism was suggested by \cite{chitta23} using \hri\ data from a coronal hole, where they reported that picoflare jets, powered by magnetic reconnection, contribute to the solar wind. There is evidence that both jetlets and picoflare jets originate from mixed-polarity solar locations, as do the larger-scale coronal jets.  
While these remain possibilities, here we present a different, non-coronal-jet-like candidate for powering the corona and solar wind. 

\begin{figure*}
	\centering
	\includegraphics[width=\linewidth]{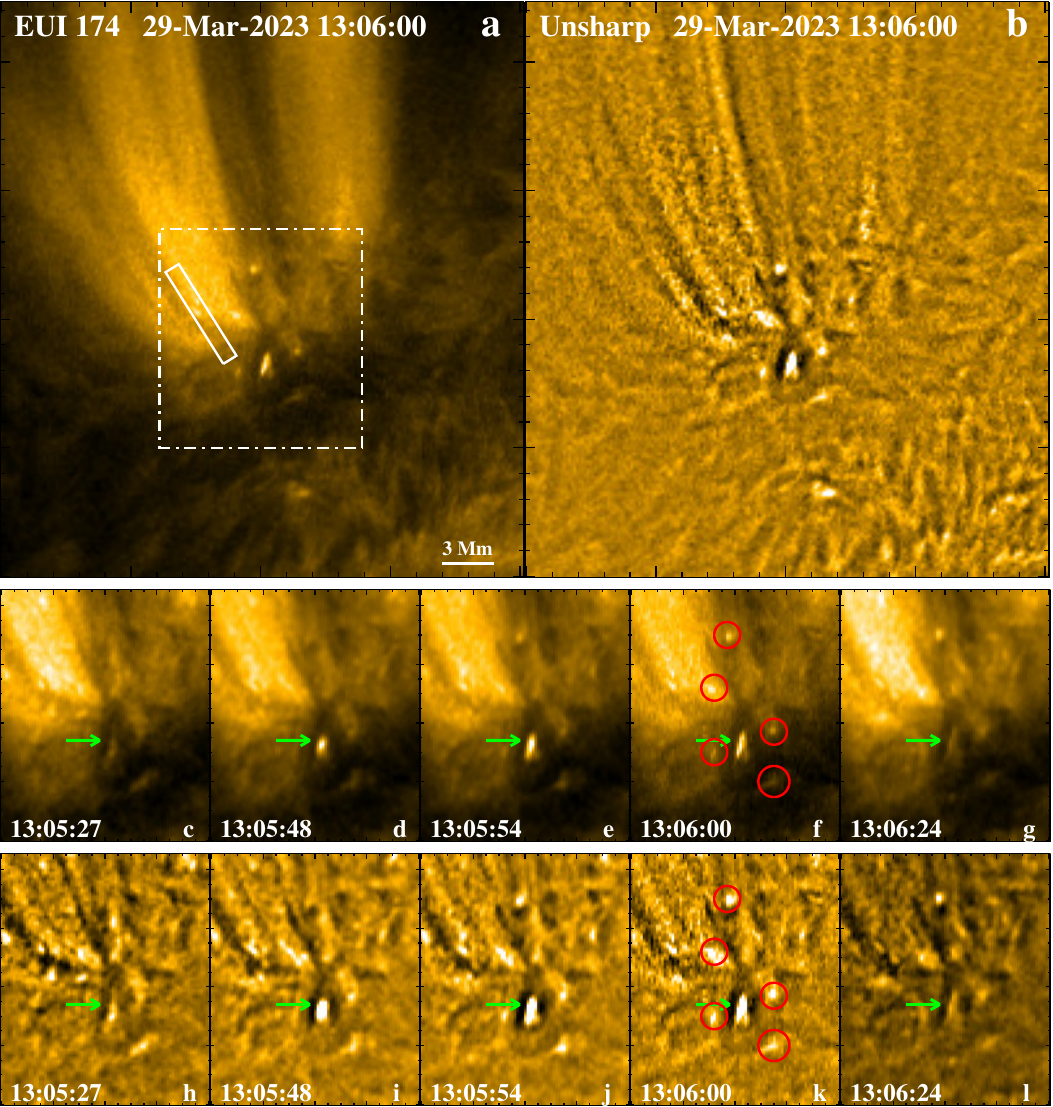}
	\caption{Examples of network microflashes at a plume base. Panel (a) shows an \hri\ 174 \AA\ image of a solar plume and several microflashes. The dotted-dashed white box in (a) outlines the field of view (FOV) displayed in panels (c--l). Panel (b) shows an unsharp mask version of the same image.  Panels (c--g) and (h--l) display \hri\ 174 \AA\ images and unsharp masked images, respectively, of network microflashes in the plume base. The green arrow points to the evolution of a non-plume-base network microflash throughout its life. The red circles encloses other microflashes that are visible in the \hri\ frames. The white diagonal slit in (a) shows the path of the time-distance plot of Figure \ref{fig6}a. The animation (Movie1) runs from 12:40 to 13:15 UT. The animation is unannotated and the FOV is same as in Panel (a). 
	}
	\label{fig3}
\end{figure*} 

We report fine-scale transient brightenings at the base of plumes, in very high spatial (142 km) and temporal (3 s) resolution 174 \AA\ coronal EUV images from Solar Orbiter's \citep{muller2020} Extreme Ultravoilet Imager (EUI)/high-resolution imager (\hri\ \citealt{rochus2020}). These brightenings  appear in  network clumps of unipolar magnetic flux at the base of plumes.  We call these events `network microflashes' (Figure \ref{fig3}).  Network microflashes can also be seen in non-plume-base unipolar network magnetic flux, but only at about 10\% of their number density found in the network flux hosting a plume. This paper proposes that network microflashes are made by bursts of reconnection that possibly power the corona and solar wind along the open magnetic field stemming from the network. 


\begin{figure*}
	\centering
	\includegraphics[width=0.77\linewidth]{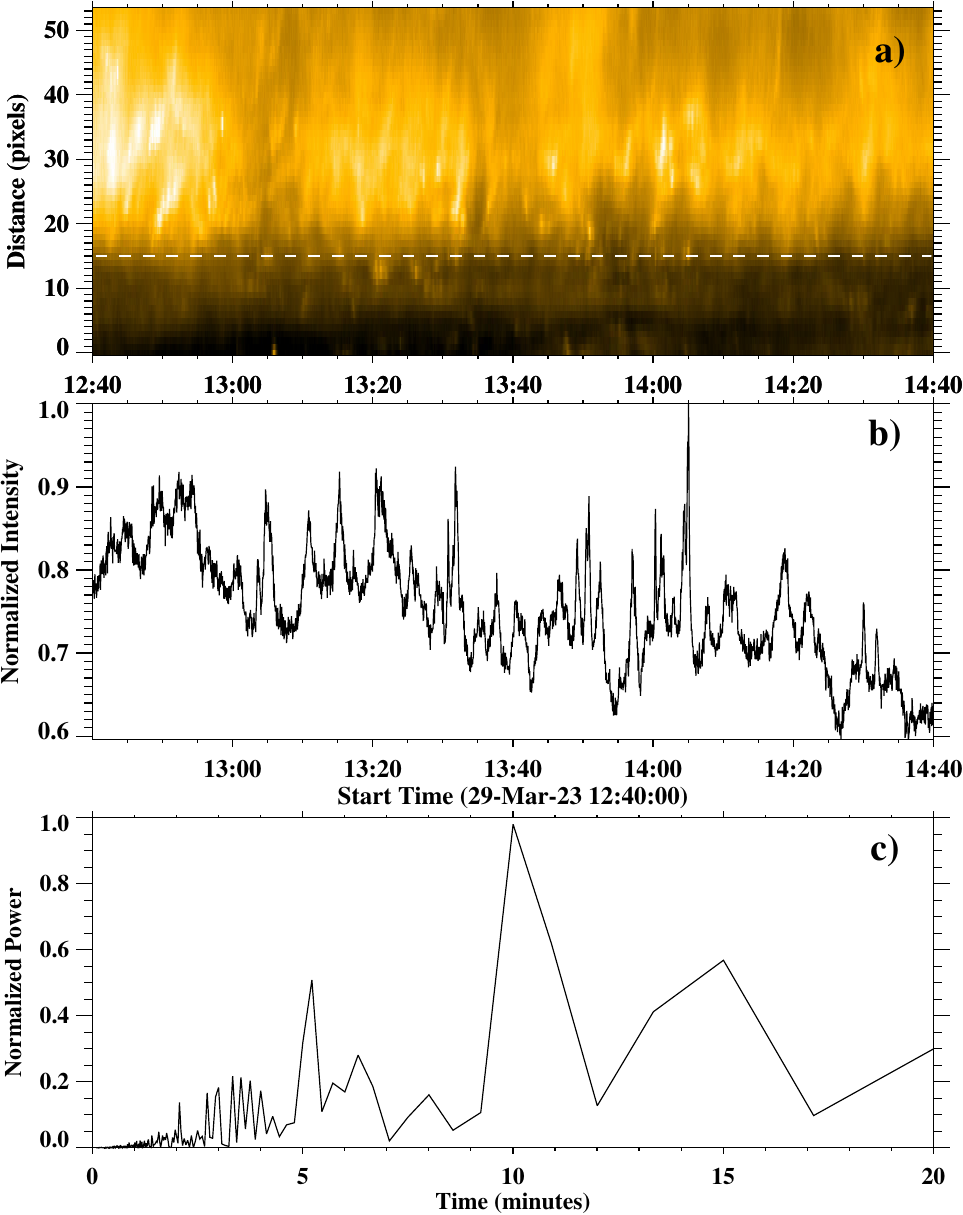}
	\caption{Examination of the periodicity of network microflashes. Panel (a) shows the \hri\ 174 \AA\ total intensity time-distance map plotted along the rectangular slit of Figure \ref{fig3}a. The right-leaning bright structures are the microflashes and their upflows that appear within that slit. Panel (b) shows the \hri\ 174 \AA\ intensity time profile from the white-dashed line in panel (a). The time profile shows peaks at intervals of 5--15 minutes. Panel (c) shows the power spectrum of the intensity time profile in (b). } \label{fig6}
\end{figure*} 

\section{DATA, Methods, and An MHD Model}\label{data} 

\subsection{Data}

To study network microflashes, we use high spatial resolution (pixel width  0.492"; \cite{rochus2020}) and high temporal cadence (from 3 to 10 seconds) images from EUI's \hri. For our analysis, we use \hri\ Level 2 data\footnote{https://www.sidc.be/EUI/data/L2/} from Data Release 6 \citep{euidatarelease6} from two days: 29-March-2023 and 26-October-2023. Solar Orbiter observed the Sun at 0.394AU on 29-March-2023, with a pixel size  of 142 km at 3 sec of cadence. On   26-Oct-2023, it observed the Sun at 0.488AU, with a pixel size of 172 km  at 10 sec of temporal cadence. There were two hours of \hri\ data coverage on each day. \hri\ provides images in 174\AA\ passband, in which  \FeIX\ and \FeX\  are primary emission lines characterizing the coronal plasma of $\le$1 MK.  The separation angle between Solar Orbiter and earth was 2.5\degree\ on 29-Mar-2023 and 28\degree\ on 26-Oct-2023.


For our analysis, we also use EUV images from the Atmospheric Imaging Assembly (AIA; \cite{lem12}) onboard Solar Dynamics Observatory (SDO). AIA takes images in seven different EUV channels with 0.6" pixels at 12 s cadence. We mainly use AIA 171 \AA\ images  because plumes are  best seen in this channel, and because the AIA 171 \AA\ is the closest AIA channel to the \hri\ 174 \AA\ passband. AIA 171 \AA\ forms around 0.65 MK and is centered on  an \FeIX\ line.  Network microflashes are barely visible in AIA images. Therefore, we  use AIA 171 \AA\ only for context (Figure \ref{fig1}). We use line-of-sight magnetograms from SDO/Helioseismic and Magnetic Imager (HMI; \cite{scherrer12}) to show  the network magnetic flux in and near the base of the plumes, and around the base of the microflashes. 

\subsection{Methods}\label{method}

To enhance the visibility of the \hri\ network microflashes, we applied an unsharp masking technique. To do the unsharp masking,  we first smooth the \hri\  images (10 $\times$ 10 pixels) and then subtract the smoothed images from the corresponding original (unsmoothed) images. These unsharp-masked images are used  to help display the network microflashes in some of the figures.  For the manual measurements, each microflash's  speed, duration, length, and width are measured from the original images.  The extension speeds of manually selected network microflashes are measured  from time-distance maps, from the slope of  the time-distance track. The length is measured  along the longer extent of the network microflash whereas the measured  width is the cross-sectional width of the network microflash. The length and width measurements are both done during the peak brightness of the microflash. The lifetime of microflashes is obtained by following them from when they start to turn on until they completely fade away. Additionally, for the manually selected microflashes, we also measured  the lifetime from the time-distance maps. 

\subsection{Automated detection and measurement of microflashes}

We implemented an automated microflash detection method that uses a Difference of Gaussian (DoG) approach to identify microflashes. In this method, images are first blurred with two Gaussian filters of different standard deviation widths. Subtraction of the two blurred images enhances features whose size corresponds to the range between the widths of the Gaussian filters.  This method assumes microflashes are bright features on a dark background, so we subtract the mean of a quiet region before applying the detection routine. Before applying the detection routine, we also perform 2 $\times$ 2 pixel unsharp masking on the original EUI images and constrain our search to the white dashed box shown in Figure \ref{fig7}a. For our detection threshold, we require the microflashes to have an intensity greater than the mean of the quiet region plus 3 standard deviations of the mean of the quiet region. For this exercise,  we take the quiet region to be a dark region where there is no significant transient brightening. Once a microflash is detected in an image, the method finds whether the microflash persists in subsequent images. For a microflash in the next image to be identified as the continuation of the microflash in the previous image, it must be centered within 2 pixels of the previous microflash’s center. That is, the maximum acceptable plane-of-sky migration speed for a microflash is $\approx$ 95 \kms (= 2$\times$ 142 km/3s). If more than one detected microflash fits this criterion, the one that is the least distance from the previous image’s microflash is chosen. This process continues until no further matches are found. Using this routine, we identify approximately 6000 microflashes in two hours. Microflashes that appear in only a single image are not counted.

Comparison with our manual detection and measurement of microflashes shows that while the 3-sigma detection threshold identifies the correct number of microflashes, it underestimates the microflash lifetime. To address this, we conduct a second search around our 3-sigma grouped microflashes with a 2-sigma detection threshold. This above-2-sigma detection uses the same method as the above-3-sigma detection scheme, but now it is restricted to a 2-pixel radius around the start and end positions of the microflashes. We continue to search at this lower threshold around a microflash until no detection can be matched. 
To measure the size of these microflashes, we perform 2D Gaussian fitting to a microflash at its peak brightness. We consider a 20 $\times$ 20 pixel box centered on the microflash and normalize the data by dividing with that field of view’s median value for each time in the observation. From our manual size measurements, we know these features to be small (length/width a few pixels) which informs the standard deviation bounds we use for Gaussian fitting. This 2D Gaussian fitting does not work perfectly for our entire sample of microflashes (especially those that are very small or on bright backgrounds), but as a statistical approximation, it is a useful characterization of the overall size and shape of the detected microflashes.

We obtain the speed of each microflash from the images that have 2 $\times$ 2 unsharp masking.  We estimate the speed from the displacement of the microflash centroid between its first detection and peak time in the original images. Each measurement is derived from images processed with 2 $\times$ 2 unsharp masking. We also applied this technique to detect microflashes in a nearby non-plume network region, finding that the number density of microflashes in this region is quite low, only about 10\% of the number density of microflashes in a plume-base network  flux clump. 

\subsection{Data-driven Global 3D MHD Model}
To calculate the magnetic field into the solar atmosphere and farther into the heliosphere, we use the Alfv\'en Wave Solar Model (AWSoM), which is a data-driven global MHD model within Space Weather Modeling Framework \citep{toth2012}. In particular, AWSoM uses synoptic or synchronic photospheric magnetograms (e.g., from HMI/SDO) as lower boundary conditions for the magnetic field, thereby constraining the large-scale coronal and heliospheric structure with real solar data. We will use the AWSoM code primarily for modeling the magnetic field out into the heliosphere.  An advantage of this code over a simple Potential Field Source Surface (PFSS) field extrapolation is that the additional pressure resulting from the dynamic and thermodynamic inputs in the AWSoM model better represent the field than would the PFSS alone, specifically in regions of the heliosphere where the plasma beta is high \citep{gary01}.  We will not attempt to simulate the physical consequences of inputting microflashes into the AWSoM code, which is beyond the scope of the current investigation.  For completeness however, in the following we provide details of the AWSoM code.

The simulation domain extends from the upper chromosphere to the corona and heliosphere beyond Mars' orbit \citep{holst2014}. The initial inner boundary conditions for electron and proton temperatures Te and Ti and number density n are  Te = Ti = 50,000 K and n = 2$\times$10$^{17}$ m$^{-3}$, respectively. This density at the inner boundary allows chromospheric evaporation to self-consistently populate the corona with an appropriately high density, as found in the Sun's atmosphere. Moderately changing the initial inner boundary density and temperature does not otherwise have a significant influence on the global solution \citep{lionello2009}. The inner boundary  magnetic field is specified with full-surface magnetic maps. In this study, we used the synchronic magnetic maps based on the Lockheed Martin surface flux transport model \citep{schrijver2003}. The initial conditions for the solar wind plasma are specified by the Parker solution \citep{parker58}, while the initial magnetic field is the PFSS model field obtained with the Finite Difference Iterative Potential Solver (FDIPS; \cite{toth2011}). The steady state solar wind solution is obtained with a local time stepping and second-order shock-capturing scheme \citep{toth2012}.

Aflv\'en waves are driven at the inner boundary with a Poynting flux that scales with the surface magnetic field. The solar wind is heated by specified Alfv\'en wave dissipation and accelerated by thermal and Aflv\'en wave pressure. AWSoM uses a phenomenological treatment of Aflv\'en wave dissipation, in which the wave spectrum is not resolved. Instead, the total energy densities of the counter-propagating waves are calculated. Electron heat conduction (both collisional and collisionless) and radiative cooling are also included in the model. In addition, the electron and proton temperatures are treated separately, with the two species being coupled by collisions.  By incorporating a physically consistent treatment of wave reflection, dissipation, and heat partitioning between the electrons and protons, the AWSoM reproduces reliable realistic global solar corona conditions \cite[e.g.,][]{sokolov2013,jin2013,jin2017,sachdeva2019}.


\section{Results} \label{sec:res}
	\subsection{Network Microflashes Concentrated at the base of Plumes}
	
Figure \ref{fig3}a is from the 29 March 2023 data set, and shows an \hri\ image of a plume bush containing many fine-scale brightenings (network microflashes) at its base. The microflashes  are better visible in  unsharp masked images (Figure \ref{fig3}b, h--l, and Movie1).  Some of the bigger network microflashes are barely noticeable in lower-resolution Solar Dynamics Observatory's (SDO) AIA 171 \AA\ images. In the Appendix (Figure \ref{A1}), we show an AIA 171 \AA\ image of the same plumes near the time of the \hri\ image. AIA 171 \AA\ images blur out the microflashes seen in \hri\ 174 \AA\ images.  SDO/HMI magnetograms show that these plumes are rooted in predominantly unipolar (negative-polarity) magnetic flux clumps (Figure \ref{A1}d). Typical microflashes occurring in the base of the plume at a given time are visible in Figure \ref{fig3}f,k (see two upper circles). Some microflashes appear away from the plume base (lower circle in panels (f and k)). It is important to note that all microflashes are rooted in the magnetic network flux: most occur in the cores of network flux concentrations, while only about 10\% are rooted in relatively weaker network flux clumps with no obvious plume structures. Some microflashes appear slightly elongated along the open field, e.g. one clear example is pointed out by the green arrows. Other microflashes appear nearly circular in shape (e.g. microflash inside the upper red circle of Figure \ref{fig3}f). This could be either because they are truly circular, or because their extension part is very dim, or because they lengthen along the line of sight. Microflashes often occur multiple times at the same location. Our analysis indicates that microflashes occurring at the same location do not exhibit a regular cadence, but rather appear at irregular intervals. In some cases, repeated events are observed at the same location with varying time separations, for example on the order of a couple of minutes ($\sim$2 minutes) to around 10 minutes. 

Figure \ref{fig6}a displays numerous bright microflashes that appear within the slit shown in Figure  \ref{fig3}a during the two hours. In order to examine any periodicity of these microflashes, we plotted intensity along the white dashed line in Figure \ref{fig6}a. The plot in Figure \ref{fig6}b shows clear intensity spikes (or microflashes) that occur roughly 5--15 minutes apart. Further, we computed the power spectrum of the intensity plot of Figure \ref{fig6}b, and show it in Figure \ref{fig6}c. The power spectrum has peaks near 5, 10, and 15 minutes. This is in agreement with our wavelet analysis that also shows high power at  $\sim$ 5, 10, and 15  minutes (Appendix, Figure \ref{A2}). The intensity-time plots at other locations in the plume base show microflashes roughly  5-15 minutes apart. These findings suggest that microflashes are often triggered by photospheric 5-minute p-mode oscillations \citep{inogamov-p-mode96}. Such oscillations are already strongly suspected  to play a role in triggering small-scale eruptive events \citep[e.g.][]{ning2004,doyle2006,gupta15,sterling20,kumar22}. Therefore, the detection of a 5-minute periodicity in microflashes is consistent with these earlier investigations.

Moreover, we speculate that the preparation time for a microflash may, in some cases, exceed five minutes; therefore, a microflash would tend to form every other p-mode oscillation, rather than every p-mode oscillation, and this potentially explains the power-spectrum peak near 10 min.  Likewise, there would be a peak at 15 minutes because sometimes the preparation requires more than ten minutes. 

	Similar periodicities of ~10–15 minutes have also been reported using Solar Orbiter's \hri\ data \citep[e.g.][]{weitz2025,baweja2025}. They have also been reported before in several studies using Solar Ultraviolet Measurements of Emitted Radiation (SUMER; \citealt{Wilhelm-sumar1995}) and AIA data in coronal loops and plumes \citep[e.g.][]{banerjee2011,prasad2015,kumar22}. The cause of these is not well understood. In coronal plumes, the observed ~10–15 minute periodicities may arise from multiple processes. For example, inclined magnetic fields lower the acoustic cutoff frequency, enabling leakage of long-period magnetoacoustic waves from the lower atmosphere into the corona \citep{Bel1977, pontieu2004,pontieu2005}. Alternatively, such periodicities could result from quasi-periodic reconnection at plume footpoints \citep{DeMoortel2009, prasad2015}. Distinguishing between wave leakage and reconnection-driven flows remains challenging, but both mechanisms plausibly could  contribute to the observed periods.

\begin{figure*}
	\centering
	\includegraphics[width=\linewidth]{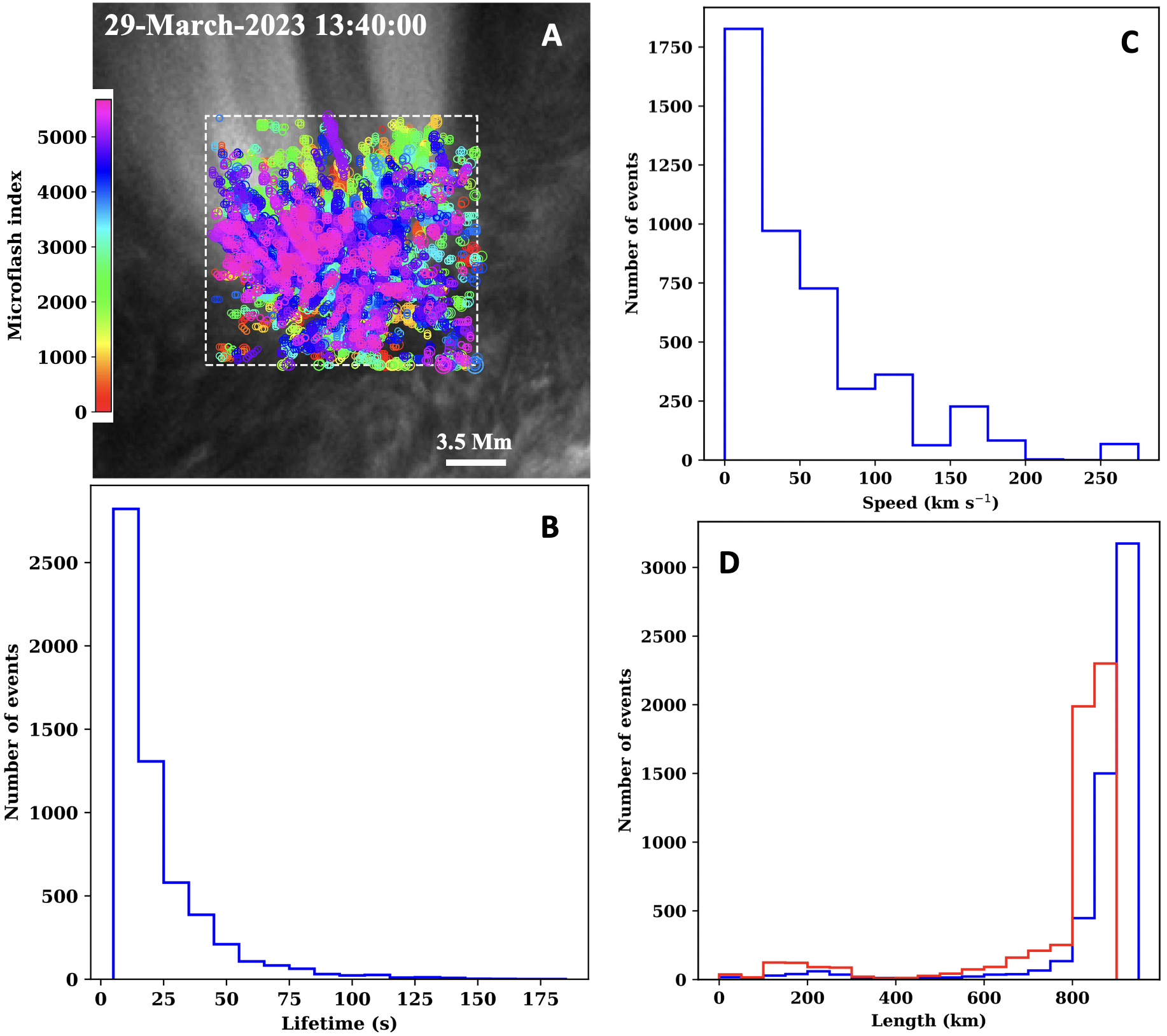}
	\caption{Automated detection of network microflashes. The different colored circles in panel (a) mark different microflashes detected over the entire \hri\ time span. The white dashed region is the FOV searched for microflashes. The background image is the \hri\ image at13:40:00. Each microflash is indexed (1– $\sim$6000) by its time of appearance, with color indicating detection time from red (earliest) to pink (latest), in accord with the corresponding color bar in (a). Panels (b--d) show histograms of microflash speed, lifetime, length (blue) and width (red), respectively, from the automated detection method. The mean values of the speed, lifetime, length and width are 50 $\pm$ 50 \kms, 22 $\pm$ 20 seconds, 860 $\pm$ 150 km and 770 $\pm$ 200 km, respectively. } \label{fig7}
\end{figure*} 

\subsubsection{Using the Automated Detection Method for a Larger Sample}\label{auto}

To measure microflash properties, we applied an automated microflash detection method (Section \ref{method}) to the 29-March-2023 two-hour sequence of \hri\ 174 \AA\ images in the plume-base field of view (FOV) of the white dashed box of Figure \ref{fig7}a. Each nest or string of same-color circles in that box marks a detected microflash. The different colors help in distinguishing nests or strings that overlap. The extent of most colored nests/strings in Figure  \ref{fig7}a is much smaller than the extents of typical network jetlets. From using this method, we found a total of about 6000 microflashes at 3-$\sigma$ above background intensity (Section \ref{method}), giving about one new microflash beginning every second in that region. We do not count microflashes that  appear in only a single image. Each counted microflash lasts at least 3s. 

The microflashes moves with an average speed of 50 $\pm$ 50 \kms. We note that the distribution of speeds (in Figure  \ref{fig7}c) is skewed toward lower values, with the most common values in the 0--25 \kms\ bin.  On the other hand, these speeds are measured in projection against the disk. Since the region is near disk center, and if most motions are approximately radial, the true speeds are likely higher than those shown in Figure  \ref{fig7}c.  Given these considerations, 50 \kms\ is a reasonable estimate for typical microflash speeds, although the scatter is large.  We obtain a mean microflash lifetime of 22 $\pm$ 20 seconds (Figure \ref{fig7}b). As discussed for the speeds, the lifetimes are also not normally distributed, and most values are clustered in the 3--12s bin. The mean microflash length and width came out to be  of 860 $\pm$ 150 km and 770 $\pm$ 200 km, respectively. 
We verified the findings of our automated method by manually measuring these aspects of 50 manually selected microflashes (Section \ref{manual}). The measurements obtained by the automated method are in a reasonably good agreement with those obtained from the manually measured microflashes.  A caveat is that the manual detections focus on a limited sample of 50 isolated and prominent events, whereas the automated selection depends sensitively on the chosen intensity threshold. As a result, the number of identified microflashes can vary significantly with the adopted threshold.

The average lifetime of our flashes is shorter than that of EUI microjets \citep[5 minutes;][]{hou21} and campfires \citep[10 minutes;][]{panesar2021}. Their lifetime is comparable to that of EUV bright dots observed by the high-resolution coronal imager (Hi-C) in active regions \citep[$\sim$25 s;][]{regnier2014,subramanian2018} and \hri\ bright dots \citep[50 s;][]{tiwari22}. Many of these flashes exhibit speeds similar to those of spicules \citep{pontieu-pasj07}. Additionally, their observed lengths are much smaller than those of EUI microjets \citep[7700 km;][]{hou21} and campfires \citep[5400 km;][]{panesar2021}. 

\begin{figure*}
	\centering
	\includegraphics[width=\linewidth]{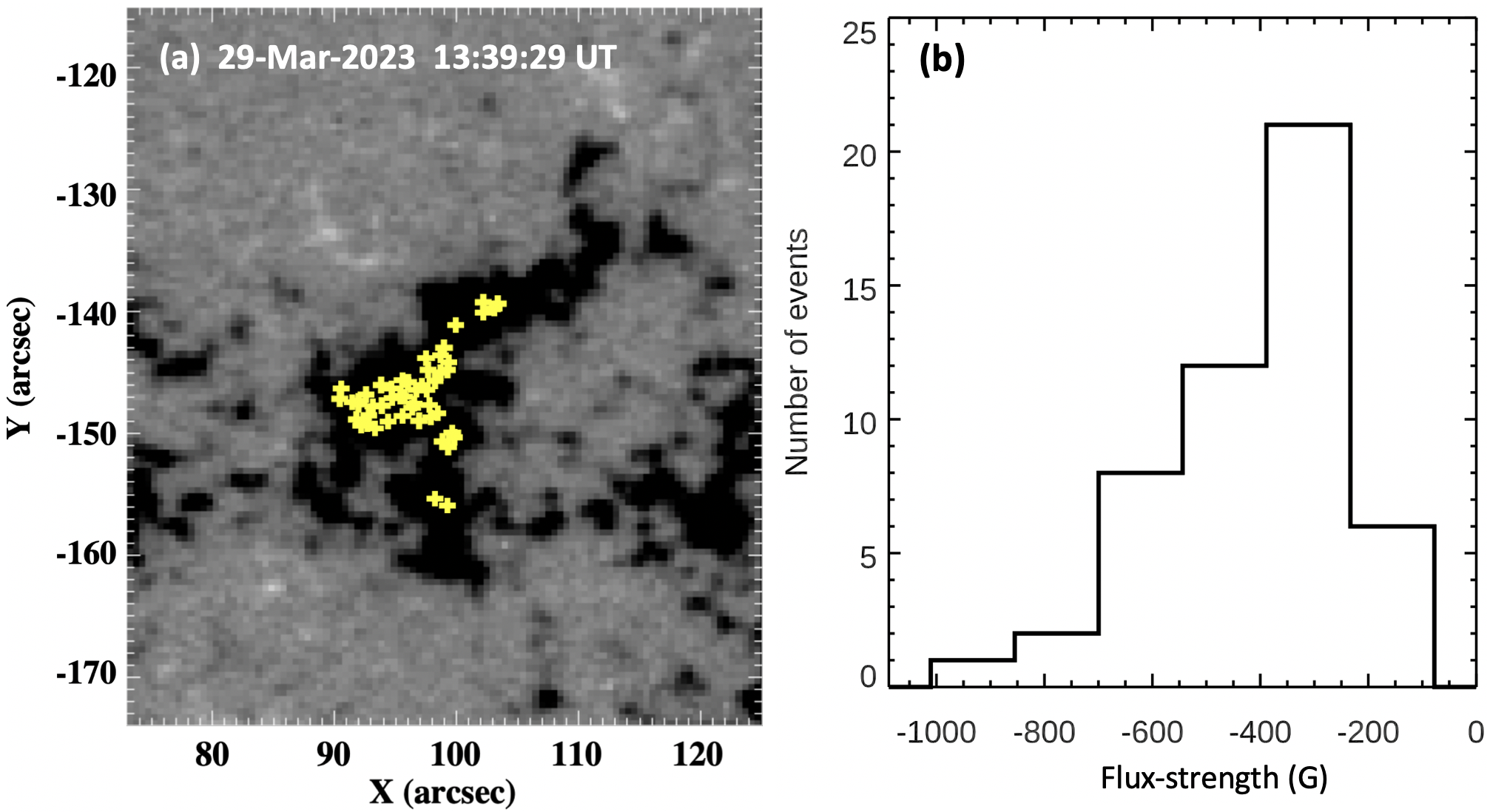}
	\caption{Magnetic Setting and Flux-strength  histogram of 50 random manually selected microflashes. (a) Locations of the peak-brightness centroids of the 50 microflashes, marked by yellow ``+” symbols, overlaid on the HMI plume-base flux map. (b) Histogram of HMI flux-strengths measured at the 50 microflash centroids shown in panel (a).} \label{flux}
\end{figure*} 

\subsubsection{Magnetic Setting}\label{mag}

While the microflashes appear to be rooted in visibly unipolar magnetic flux, we examined HMI magnetograms in detail  for evidence of opposite-polarity (minority) magnetic flux at the microflash sites. For this, we used 50 random manually selected microflashes (Appendix \ref{manual}). In Figure \ref{flux}, we show the magnetic flux locations and strengths of the 50  microflashes.  Figure \ref{flux}a shows the locations of the peak-brightness centroids of the 50  microflashes within the plume-base flux. The microflashes are concentrated well inside the enhanced network flux patch at and near the plume base, suggesting that they preferentially occur in the patch's stronger magnetic flux.

Figure \ref{flux}b is the histogram of the HMI-magnetogram flux strength at the 50 microflash centroids. The histogram shows that a great majority of the microflashes (44 out of 50) are  in negative flux stronger than 200 G and a clear majority (35 out of 50) are in negative flux stronger than 300 G. While this result does not prove that most plume-base microflashes are unipolar (i.e., not located at unresolved polarity inversion lines), it is quantitative circumstantial evidence favoring that most microflashes are in truly unipolar magnetic flux. That is, favoring that microflashes are not rooted at places in the HMI magnetogram where the majority-polarity flux strength has a local depression due to undetected embedded minority-polarity flux.

A network field strength exceeding 200 G does not preclude the presence of opposite-polarity flux, as discussed by \cite{wang-y-m2016} and \cite{wang2020}. These studies \citep{wang-y-m2016,wang2020} show that small-scale opposite-polarity flux and associated small-scale loop structures (as seen in EUV images) can exist even within apparently unipolar network concentrations, and may play an important role in plume dynamics. As just noted, our histogram results do not  rule out the presence of such small-scale opposite-polarity flux at the plume base.
	

We can, however, provide two additional arguments that, although circumstantial, do support that the microflashes we are discussing here do not originate at the site of undetected mixed-polarity magnetic regions. 
	
	First of all, collimated ejections such as those that produce jetlets often result from eruptions at sites of flux cancelation at the feet of magnetic field lines that are rooted at the edge of otherwise apparently unipolar network flux clumps and reach into the corona \citep[e.g.][]{panesar18b}.   In our observations, however, we detect very few such collimated ejections; instead, the microflashes appear significantly fainter and smaller than the previously reported collimated outflows above unipolar plages. Moreover, the spires of jetlets are rooted in bright bases, and our microflashes appear without such bright bases.  This is consistent with the microflashes having an origin different from that of jetlets, and thus not resulting from flux cancelation on an undetected scale. 

Our second argument is based on the fact that among the strongest evidence for hidden multipolar fields in seemingly unipolar magnetograms is the detection of compact loops in EUV  \citep{wang-y-m2022}. In contrast, we detect no clear evidence of such small-scale loop structures in our microflash feet in either \hri\ 174 \AA\ and AIA 171 \AA\ images.  While it is possible that such loops are present but obscured by the overlying plume emission (``plume bushes"), we emphasize that this lack of EUV-small-loop observations consistent with the microflashes having a unipolar origin.   Based on our observations and the arguments presented here, we will suggest a possible unipolar mechanism for producing microflashes in Section \ref{DrivingMech} below.

\subsubsection{Network Microflashes Production Rate, Mass Flux, Magnetic, Thermal and Kinetic Energies} 

An upper bound on the magnetic energy released by a single microflash (B$^2$$\times$V/8$\pi$) is estimated to be  2.0 $\times$ 10$^{26}$ erg. For this estimate, the magnetic field low above the network is taken to be 100 G. We note that magnetic field strengths can vary significantly within network regions across different parts of the Sun. Observationally, field strengths in plume-base regions range from approximately 200 to 600 G and more \citep{avallone18}. Furthermore, nonlinear force-free field extrapolations have indicated magnetic field strengths of 500 G and higher within active regions in low lying coronal loops  \citep{thalmann14}. In  our plumes, we measure average field strength to be $\sim$100 G.  Because there is considerable uncertainty in the above mentioned measurements, we take 100 G as an approximate representative value. Moreover, our histogram of Figure \ref{flux}b  indicates that, if anything, our 100 G estimate is an (\textit{underestimate}) of the flux strength at the base of most of the microflashes.  Not all of the flux would go into powering a microflash, but this does support that our estimate of 100 G in the energy calculation is reasonable.  

For the volume (V), we used the average microflash length and width (V = l$\times$w$^{2}$). This magnetic energy is in the range of that of campfires \cite{panesar2021}, coronal hole jets \citep{pucci13}, and coronal bright points \citep{priest94}. This value is about an order-of-magnitude lower than that for active region jets \citep{sterling17,panesar2025}, active region braided loops \citep{cirtain13} and subflares \citep{tiwari14}. The size of microflashes is  smaller than most previously known transients, e.g. fine-scale loops, dots, jets/surges, and campfires \citep{tiwari19,berghmans2021,panesar23,nobrega2025}. 

 The estimated thermal (E = 1.5N$_{e}$k$_{B}$TV) and kinetic energies (0.5N$_{e}$m$_{p}$V$v^{2}$) of an average network microflash   are $\sim$ 10$^{24}$ erg, and $\sim$ 10$^{23}$ erg, respectively. Here, N$_{e}$ is electron number density, k$_{B}$ is Boltzmann constant, T is temperature, V is volume, m$_{p}$ is mass of proton and $v$ is the  speed of a microflash. We have taken an average  speed ($v$) of 50 \kms, derived from our observations (Figure \ref{fig7}), an electron density of 10$^{10}$ cm$^{-3}$ \citep{withbroe77,young-elecdenisty2009,cirtain13}, and a microflash plasma temperature  of 10$^{6}$ K in agreement with the sensitivity of the 174 \AA\  channel to emission from $\approx$1 MK plasma \citep{rochus2020}. 
 
 The estimated kinetic energy is of order of that of a coronal EUI microjet's kinetic energy \cite[$\sim$10$^{23}$ erg;][]{hou21} and  two order higher than picoflare jet's kinetic energy \citep[$\sim$10$^{21}$ erg;][]{chitta23}. The thermal energy of  a network microflash is of the order that is estimated (10$^{24}$ erg) \citep{parker88} for nanoflares. If we assume that during the microflash-making burst of reconnection about 2\% of the upper-bound  magnetic energy for a microflash gets converted into Alfv\'en waves that propagate out along reconnected open field, then the magnetic energy released during a microflash's reconnection burst would be $\sim$ 4 $\times$ 10$^{24}$ erg. According to \cite{yokoyama95},  the ratio of Alfv\'en wave energy to the total released magnetic energy is approximately 3\%. Here we will use the more conservative 2\% value for our estimate. 
 Considering the expansion of the plume area at its top, say the radius expansion is a factor of two, giving 4 $\times$ 10$^{18}$  $cm^{2}$ for a plume top's area, network microflashes mark release of sufficient energy to power the solar corona and solar wind in plumes in coronal holes ($\sim$ 10$^{6}$ erg cm$^{-2}$ s$^{-1}$) \citep{withbroe77}.      

 The mass-loss rate \.m from network microflashes can be estimated using the formula: \.m = 4$\pi R^{2} \rho v f_{CH} f_{PL} f_{NF}$, where R is the radius of the Sun, m$_{p}$ is the proton mass, $\rho$ = m$_{p}$10$^{10} ~gm~cm^{-3}$ is the microflash mass density, $v$ = 50 \kms\ is the measured average speed of network microflashes, $f_{CH} \sim 0.1$ is the fractional area roughly covered by coronal holes on the Sun \citep{harvey02}, $f_{PL} \sim 0.1$ is the fractional area covered by plumes in coronal holes \citep{ahmad77}, and $f_{NF} \sim 0.1$ is the plume-base fractional area of network microflashes in plumes at a given time. Using these values and assuming all of the microflash mass flux enters into the solar wind, the total mass loss rate for network microflashes comes out to be 5 $\times$ 10$^{12}$ gm  s$^{-1}$, which is the same order as  the mass loss rate for ``network jets" \citep{tian14} and ``picoflare jets" \citep{chitta23}. 
 
Further, an average microflash supplies $\sim$ 3.3 $\times$ 10$^{32}$ protons s$^{-1}$ to the solar wind, if we assume that all of the microflash’s up-flowing plasma escapes into the solar wind.  A total of 6 $\times$ 10$^{35}$ protons s$^{-1}$  is needed to sustain the solar wind \citep{steiger2000,wang2016,wang2020}, which requires $\sim$ 2 $\times$10$^{3}$ network microflashes (or $\sim$  10$^{2}$ plumes) to be present on the Sun at a given time to supply the solar wind’s mass loss.  For  $f_{CH} \sim 0.1$, the area of coronal holes on the Sun is $\sim$ 0.1$\times$ 4$\pi R^{2}$= 6 $\times$ 10$^{21}$ cm$^{2}$.  We estimate the area of the base of a coronal-hole plume is $\sim$  10$^{18}$ cm$^{2}$ and the area of the plume top is $\sim$  4 times that, or $\sim$  4$\times$ 10$^{18}$ cm$^{2}$.  So the top area of 100 plumes is
$\sim$ 4 $\times$ 10$^{20}$ cm$^{2}$, which agrees with the fraction of the area of coronal holes filled by plume tops being $\sim$ 0.1, as was assumed above in estimating the mass loss rate from network microflashes.

In this work, we do not claim that microflashes uniquely explain the observed fluxes, but instead propose them as a previously unrecognized candidate contributor to the solar wind. Whether they are a dominant source, a minor component among other mechanisms (such as spicules, jets, and nanoflares), or not a significant contributor at all remains an open question that requires further investigation. We emphasize that the microflashes identified here are among the smallest and most ubiquitous features observed at the base of plumes to date, and thus may represent an additional component in the overall energy and mass budget.

\begin{figure*}
	\centering
	\includegraphics[width=\linewidth]{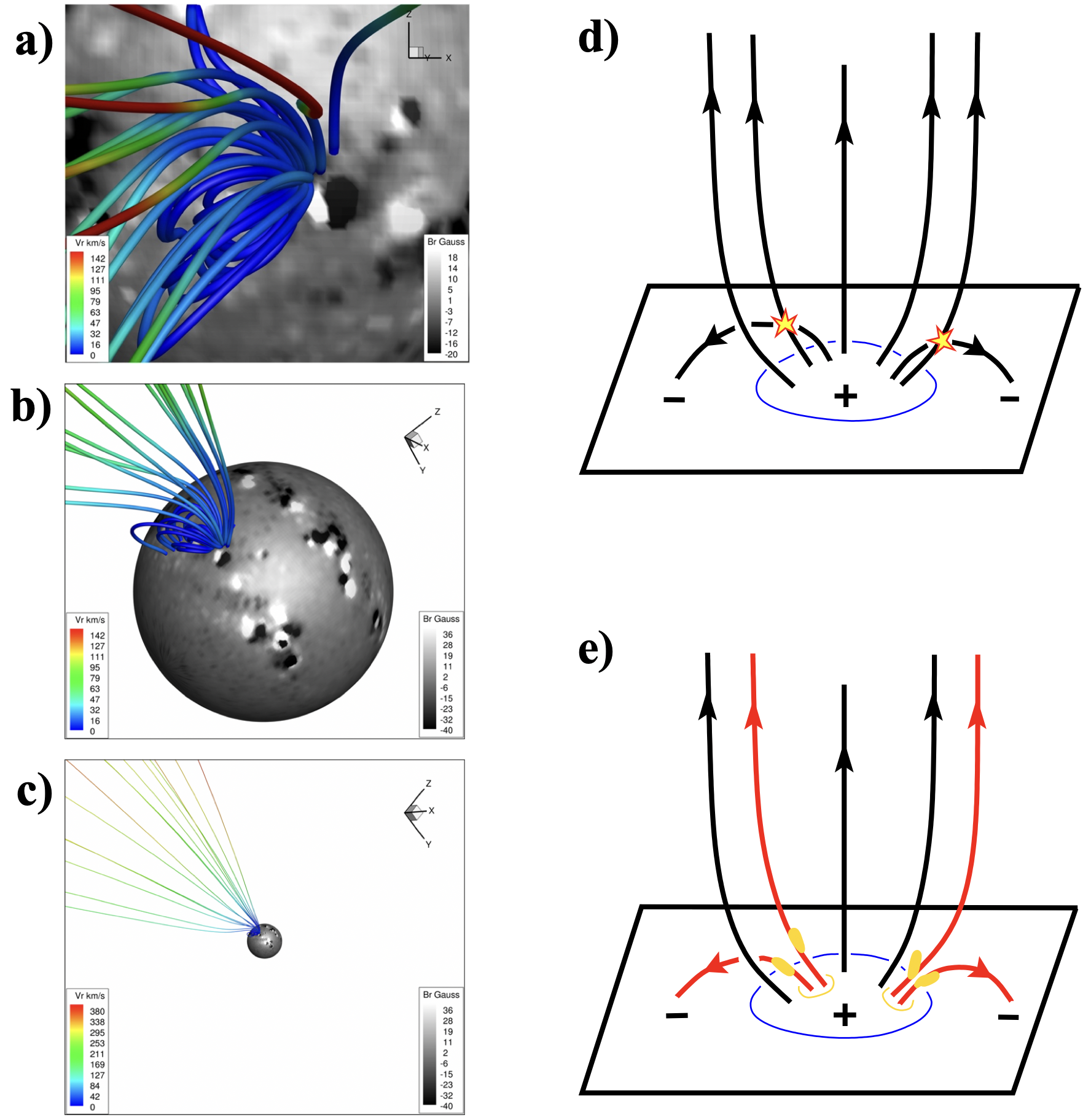}
	\caption{Models for the open magnetic field and the microflashes in the observed coronal plume. Panels (a--c) show the MHD model 3D field from the region of the 29-March-2023  plume. The radial magnetic field is shown at r = 1.006 Rs with gray scale. Selected field lines  from the plume region are shown. The color on the field lines shows the radial velocity of solar wind. Panel (a)  shows a similar point of view as the observation. Panels (b) and (c) show a different point of view. The open field lines in (b) and (c) extend to $\sim$2 and  $\sim$15 Rs, respectively. Panels (d--e) show our schematic for making unipolar \hri\ microflashes in a unipolar  magnetic network flux clump by interchange reconnection of plume-base open field with closed field from plume base to network cell interior. Panel (d) shows the start of two bursts of interchange reconnection. Panel (e) depicts the appearance of microflashes made by the bursts of interchange reconnection. The black curves are magnetic field lines before reconnection whereas the red curves are the newly reconnected magnetic field lines. Stars show the location of magnetic reconnection. The blue oval is the plume-base unipolar magnetic network flux clump. The yellow circles and solid ovals are network microflashes. The `+'  and `--' signs, respectively, are for  positive and negative magnetic polarity.} \label{skt}
\end{figure*} 

\section{Summary and Discussion}

We report on the presence of fine-scale brightenings, network microflashes,  at the base of solar coronal plumes. They are small-scale and short lived features. They appear as bright, slightly elongated, structures that extend along the plume funnels. Coaligned HMI magnetograms show that the microflashes  are rooted in the magnetic network flux lanes.  Some of the network microflashes also appear in non-plume network flux as shown in Figure \ref{fig3}. The studied examples show that all the network microflashes come from network magnetic flux, most of which have a plume bush, but some network regions do not have a plume bush. The number density of microflashes found in non-plume network regions is only about 10\% of that found at plume base network regions. Perhaps there is not sufficient magnetic field strength to form a plume in those locations \citep{avallone18}  whereas the rest of the magnetic network has all the necessary ingredients (strong enough open magnetic field) to form a plume.  Unlike jetlets \citep{panesar18b,panesar19,panesar20b} and campfires \citep{panesar2021,hou21}, \textit{network microflashes sit in evidently unipolar magnetic flux (see Section \ref{mag}).} There are only some weak minority-polarity magnetic flux elements present at the edges of magnetic flux lanes in the HMI magnetograms (see Figure \ref{A1}d). 

\subsection{Network Microflashes and their Possible Connection with Solar Wind}


To investigate the form of 3D magnetic field lines of the plume region, we performed data-driven, global magnetohydrodynamics simulations of the plume using HMI carrington maps. Figures \ref{skt}a-c show the 3D open and closed field lines  that are connected respectively with the plume and with a nearby active region from different points of view. The background magnetic field magnetogram is saturated at a low threshold in order to display weaker field regions. The MHD model shows that there are open magnetic field lines, emanating from the plume base, with closed field lines connecting to the nearby active region. The open field lines in 
Figure \ref{skt}b extend a few solar radii (Rs). Figure  \ref{skt}c displays the open field lines that extend into the heliosphere and carry solar wind of about  $\sim$400 \kms\ at 15 Rs. As the  solar wind along open field lines moves further outwards, the solar wind speed increases making it over 500 \kms\ at 20 Rs, to be in the range of fast solar wind speed. 

It is well known that coronal plumes are a major source of solar wind \citep[e.g.] []{mcIntosh10,tian11,pucci14,zangrilli20,moore23} but it is uncertain what drives the solar wind from plumes. Our observation suggest that  \hri\ microflashes are candidates for driving  fast solar wind from plumes.

Our global MHD code that we use to simulate the heliospheric magnetic field, albeit while not resolving microflash-scale physics, supports our assumption by showing  that the magnetic fields threading the observed plumes are open to the heliosphere. In our configuration of the AWsoM code, the resulting  outflow speeds along those open fields fall within the range observed speeds for the fast solar wind. We speculate a scenario in which plume-base network unipolar microflashes supply the necessary heat and momentum to the overlying open field, thereby contributing to coronal heating and the acceleration of the solar wind. 

We note that the microflashes occur at a location around which AIA images indicate there are both open and closed field lines.  To increase our confidence that the field of the microflashes is indeed open, we opt to use the AWSoM code, which more robustly models the heliospheric magnetic field than does PFSS.  The fact that the AWSoM code reproduces solar wind speeds within the observed ranges supports that its modeling of the heliospheric field is correct.  Our observed microflashes occur at the base of the AWSoM-modeled open field lines.  This supports that the microflashes occur on open field, and this in turn is consistent with the microflashes being a candidate contributor to the solar wind. Nonetheless, further investigations are necessary to confirm whether the microflashes indeed play the role suggested here.



\subsection{Network Microflashes and Other Eruptive Events}

  We do not take the microflashes discussed here to be small-scale versions of coronal jets or jetlets \cite[e.g.][]{sterling15,panesar18b,kumar22,chitta23,raouafi23} because of the following: (i)  microflashes happen in magnetic field stemming from evidently unipolar magnetic network flux unlike jets/jetlets; (ii) they do not show a broad base (or inverted Y-shaped structure) and base brightenings as jets/jetlets show during eruption; (iii) unlike jets and jetlets, most microflashes do not exhibit any spike-like extension.
Nonetheless, we cannot rule out the possibility that some of the microflashes extending along the plume direction may share similarities with jetlets and picoflare jets,  albeit they are rooted in evidently unipolar magnetic flux and do not have a relatively broader base  than the spire.

Microflashes are also apparently different from solar campfires \citep{clery2021,berghmans2021}. They are more numerous and of smaller size than campfires, but this fact alone still allows that microflashes may just be smaller, more numerous versions of campfires.  The microflashes, however, appear  without the eruption of a flux rope, while most campfires show clear evidence of an erupting flux rope at their base, both observationally \citep{panesar2021,panesar22} and in modeling \citep{chen21},  which supports that the two phenomenon are distinct features.  Furthermore, while both campfires and microflashes occur in/at network magnetic flux, to date, campfires have mainly been reported in quiet-Sun regions, particularly at mixed-polarity sites \citep{panesar2021,kahil22}, whereas microflashes predominantly occur at the base of plumes, in unipolar open magnetic field.


Microflashes also apparently differ from `plume transient bright points' \citep{raouafi14}, which typically have lifetimes of 10 minutes (in contrast to 22 s for microflashes) and are most clearly observed in the AIA 193 \AA\ channel. Plume transient bright points, however, also tend to occur at sites of visible flux cancelation, similar to jets and jetlets, but distinct from the unipolar bases of microflashes.
	

Nonetheless, some of the microflashes we report share characteristics with the Hi-C EUV bright dots observed in the 193 \AA\ channel \citep{regnier2014}, including comparable lifetimes and widths. However, the Hi-C bright dots in that study occurred within closed loops at the edge of an active region. Similarly, bright dots reported in the core of an active region and within a coronal bright point by \cite{tiwari19,tiwari22}, using Hi-C 2.1 and \hri\ 174 \AA\ data, respectively, were also inside larger closed magnetic fields. In contrast, our observations indicate that microflashes occur within open, unipolar magnetic fields at the bases of coronal plumes. It could be that some of the microflashes may occur at locations of closed loops that are too small for \hri\ to detect, in which case it is plausible that some microflashes and bright dots share a common physical origin. Therefore, we cannot exclude the possibility that some features that appear to be  microflashes may instead be  part of a broad class of small-scale dot-like and jet-like solar activity  (those exceptions might be formed, for example,  by mixed-polarity magnetic elements on too small a size scale for us to detect), and that somehow circumvent the other expectations from mixed-polarity jets that we point out at the end of Section \ref{mag}).


\subsection{Driving Mechanism of Network Microflashes}\label{DrivingMech}

Based on our \hri\ and HMI observations, we propose the schematic picture in Figures \ref{skt}d-e for the microflash driving mechanism.  In Figure \ref{skt}d, we show an open plume field rooted in a unipolar, positive-polarity network flux clump. [By ``unipolar,” we mean that all of the magnetic flux forming the plume, specifically, all flux within the blue circles in Figures \ref{skt}d and \ref{skt}e, belongs to a single polarity. This configuration differs from the typical magnetic setup of coronal jets, which involve a compact bipolar erupting flux element interacting with nearby open field \citep[see, e.g., Figure 4 of][]{panesar16b}. In the present case, unlike jets, no opposite-polarity element is observed at the plume base (within the blue circle). Moreover, the loop field that reconnects with the plume’s open field to produce the microflash does not erupt, in contrast to the jet scenario where a compact bipole first erupts and subsequently reconnects with open field to generate the jet bright point and spire \citep[see, e.g., Figure 4b of][]{panesar16b}.]
Negative flux clumps are present outside the plume base and there are loops that connect to the negative flux. 
Interchange magnetic reconnection occurs between the crossed open field and closed loops (stars in Figure \ref{skt}d).  The positive legs of some closed loops are rooted at the edge of the network flux clump and the positive legs of other closed loops are rooted in the middle of the network flux clump. The interchange reconnection results in a reconnected closed field loops (red closed field line in e) and a reconnected open field lines (red open field in e). Network microflashes appear near the positive feet of these newly-reconnected open and closed field lines.  We propose that such interchange reconnection heats network microflashes.  \cite{vasyl24} have proposed a similar reconnection scenario for making unipolar spicules.  Again, in contrast to coronal jets, our observations show no evidence of an explosive flux rope expulsion from the surface that could drive the interchange reconnection. Moreover, we do not observe small magnetic bipoles in the plume base region, which is a characteristic feature typically linked with coronal jet activity.

Although the magnetic network flux concentrations hosting microflashes appear predominantly unipolar in HMI magnetograms, it is plausible that small-scale opposite-polarity magnetic flux exists below HMI’s detection threshold \citep{wang-YM2016}. Reconnection between small loops from such unresolved minority-polarity flux and the open majority-polarity field could be responsible for the observed microflashes. However, our current observations do not show any evidence of opposite-polarity flux patches within the magnetic network flux concentrations.   Moreover, our histogram in Figure \ref{flux}b shows that the majority of our microflashes occur where the positive fiux is strong compared to the expected flux values in coronal hole locations. This means that it would require an exceptionally high-flux negative-polarity element to enter into that base region and survive and form a mixed-flux bipole that drives the microflash. Alternatively, the emerging negative-polarity flux -likely in the form of a small bipole or ephemeral region - does not persist as a distinct feature, but instead rapidly reconnects with the surrounding dominant positive flux, releasing energy along open field lines. In addition, such minority-polarity flux may already be present prior to plume formation, embedded within the converging supergranular flow field \citep{wang-y-m2022}.

Therefore, although we cannot completely rule out such a scenario due to the lack of higher-quality magnetograms for the present study, on balance, the present study supports that the microflashes are formed by a unipolar mechanism.  That being said, our study does open obvious avenues for further research.  
	Future observations with higher-resolution and higher-sensitivity instruments (such as DKIST and Solar Orbiter’s Polarimetric and Helioseismic Imager) in coordination with EUI - might possibly reveal small-scale inclusions of opposite-polarity magnetic flux at the sites of many network microflashes. Detection of such minority-polarity flux at most microflashes  would suggest that most microflashes work like conventional jets e.g. magnetic reconnection accompanying magnetic flux cancelation \citep{panesar16b,chitta17,panesar18a,tiwari19} instead of the way we propose in Figure \ref{skt}. However, in the present work, we adhere to our interpretation based on the observational evidence that we have.

Another possibility is that  network microflashes are a consequence of shock waves, without requiring magnetic reconnection. Chromospheric shock waves might produce localized brightenings in the transition region. However we did not find microflash extensions having parabolic trajectories, which are often a signature of shock waves  \citep{pontieu07} as recently reported by \cite{rui2024} in EUV bright
   tadpoles that occur at the  base of coronal loops.  Instead, our observations favor that magnetohydrodynamic (MHD) waves, generated by photospheric convective motions \citep{osterbrock61,balle11,sakurai17} (or by p-mode oscillations), propagate through the chromosphere, along open magnetic field lines and trigger the burst of magnetic interchange reconnection that drives a network microflash in our scenario (Figure \ref{skt}e). 

\section{Conclusions}

  Using Solar Orbiter (\hri) observations, we report the presence of fine-scale, short-lived  brightenings at the base of plumes. SDO/HMI magnetograms indicate that these brightenings are located in unipolar magnetic flux.  Our \hri\ observations and analysis, augmented with the magnetic field modeled out into the heliosphere with the MHD model we use, suggest that the microflashes might power  and feed the solar wind by driving plasma outflows along the open field lines. The microflash events plausibly have sufficient energy to power the corona and solar wind emanating from the network.  Our observational and simulation results further suggest that  network microflashes mark small-scale bursts of  interchange reconnection at interfaces between unipolar legs of open and closed magnetic field low above the magnetic network.

On their own the microflashes are too slow to escape into the solar wind, and so additional acceleration and heating are needed to supply the solar wind. We conjecture that this need is met by upward propagating Alfv\'en waves (e.g., as discussed in \citep{sterling2024}) formed when a burst of interchange reconnection makes a microflash. These considerations are topics for future investigations.  Here we have presented evidence that the microflashes are a possible source of the solar wind that is different from jetlets, picoflare jets, and plasma up-flows recently suggested \citep{raouafi23,chitta23,duan25} as being the source.

\begin{acknowledgments}
 We sincerely thank two anonymous reviewers for their positive and constructive comments. We acknowledge the use of  Solar Orbiter/EUI and  SDO/AIA/HMI data. AIA is an instrument onboard the Solar Dynamics Observatory, a mission for NASA’s Living With a Star program. Solar Orbiter is a space mission of international
collaboration between ESA and NASA, operated by ESA. The
EUI instrument was built by CSL, IAS, MPS, MSSL/UCL,
PMOD/WRC, ROB, LCF/IO with funding from the Belgian
Federal Science Policy Office (BELSPO/PRODEX); the
Centre National d’Etudes Spatiales (CNES); the UK Space
Agency (UKSA); the Bundesministerium für Wirtschaft und
Energie (BMWi) through the Deutsches Zentrum für Luft- und
Raumfahrt (DLR); and the Swiss Space Office (SSO).  This work was supported by
the NASA Science Mission Directorate’s Heliophysics Division by research grants from the Heliophysics Guest Investigators (HGI) program and the Heliophysics Supporting Research (HSR) program. NKP acknowledges support from NASA’s SDO/AIA (NNG04EA00C) grant, NASA's  HCSI (80NSSC25K7028) grant, and NASA’s HSR (80NSSC24K0258) grant.  SKT, RLM, VA, and NKP sincerely acknowledge support from NASA HGI grant (80NSSC21K0520), HSR grant (80NSSC23K0093) and/or NSF AAG award (no. 2307505). SKT also acknowledges support from ARC-CREST (NASA Cooperative Agreement 80NSSC23M0230). ACS and RLM acknowledge support from their
NASA HSR grant. ACS benefited from discussions at the International Space Science Instituteproject (ISSI-BJ ID 24-604) on ``Small-scale eruptions in the Sun.” This work has made use of NASA ADSABS and Solar Software. 
\end{acknowledgments}


%
%

\newpage
\appendix

\section{Data Set 1 (\hri\ and SDO Observations)}\label{ex-1}
\begin{figure*}[ht]
	\centering
	\includegraphics[width=\linewidth]{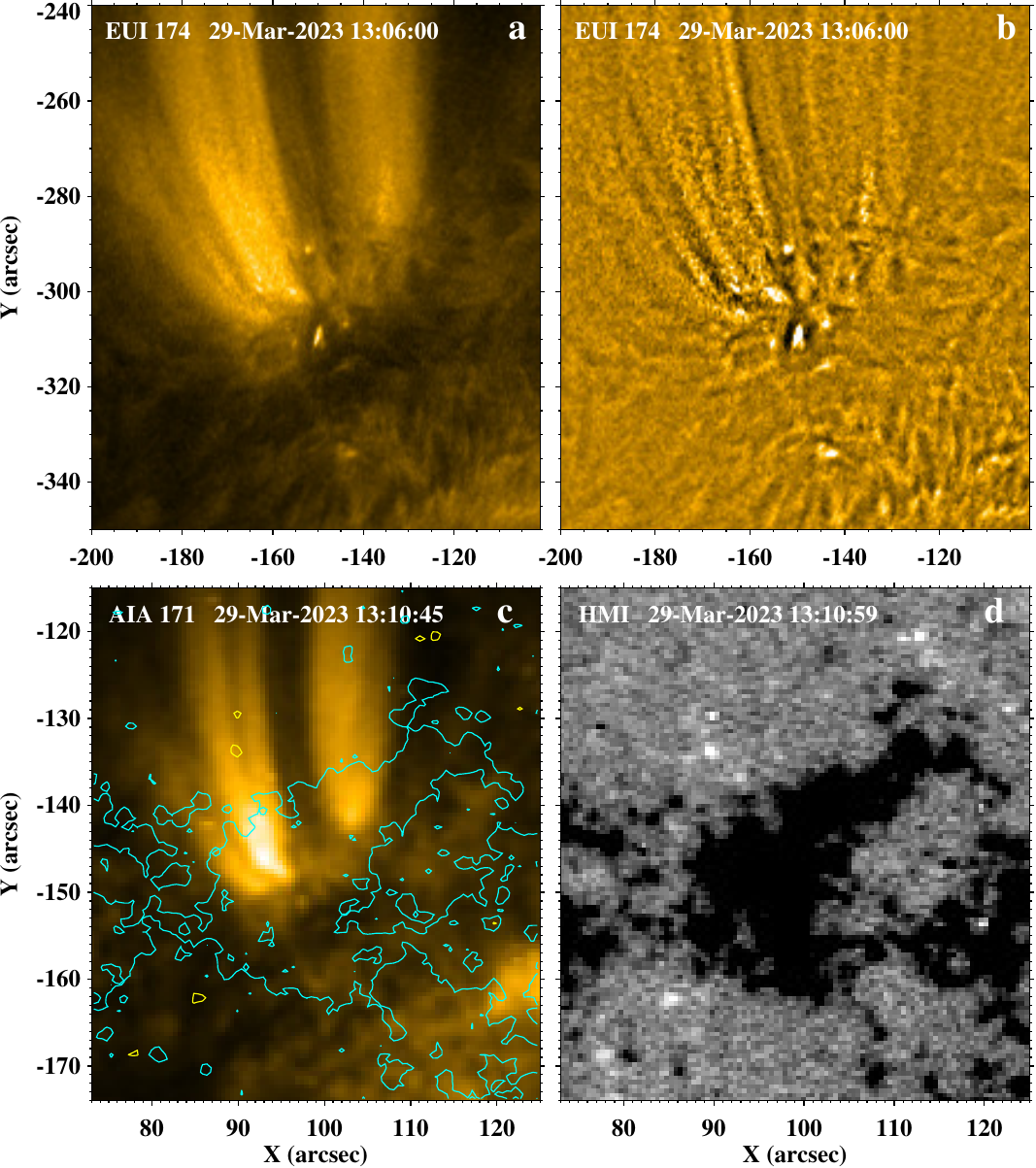} 
	\caption{The first example of network microflashes at a plume base observed by \hri\ (from Figure \ref{fig3}) and SDO.  Panel (a) shows an \hri\ 174 \AA\ image of  solar plumes and microflashes. Panels (b) shows the unsharp mask version  of the same image. Panel (c) shows the same FOV in AIA 171 \AA. Panel (d) shows the line-of-sight photospheric magnetic flux in the same FOV.  In panel (c), HMI contours, of levels $\pm$20 G, at 09:33:271 UT are overlaid, where yellow and cyan contours outline positive and negative magnetic flux, respectively.  The animation (Movie2) runs from 12:40 to 14:40 UT. The animation is unannotated and the FOV is same as in Panel (a). 
	}  \label{A1}
\end{figure*}

Figure \ref{A1} shows the same FOV that is displayed in Figure \ref{fig3}. It also includes an AIA 171 \AA\ image of the plume and its underlying photospheric magnetic field. While the high resolution images from \hri\ definitely show extremely fine-scale, transient, bright structures (microflashes) at the base of solar coronal plumes, the AIA 171 \AA\ image shows hardly any of these microflashes. The plume is  evidently rooted in the negative-polarity magnetic flux patch, with some scattered positive-polarity magnetic grains near the edges of the negative-polarity plume-base patch (Figure \ref{A1}d).

\subsection{Manual measurement of microflashes} \label{manual}

\begin{figure*}
	\centering
	\includegraphics[width=\linewidth]{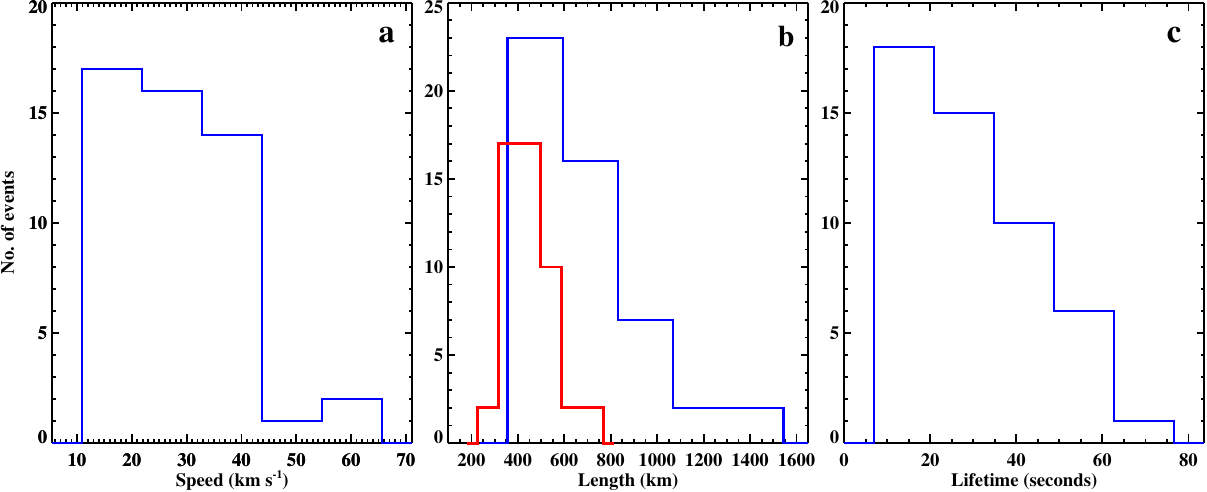}
	\caption{Histograms of the plane-of-sky speed (a), length (b), and lifetime (c) for the 50 manually selected network microflashes. The red histogram in (b) is the width of the microflashes. The bin size is different for length (235 km) and width (100 km) histograms. The mean values of the speed, length, width, and lifetime are 27.5 $\pm$ 11.5 \kms,  675 $\pm$ 250 km, 450 $\pm$ 95 km, and 30 $\pm$	15 seconds, respectively.} \label{fig5}
\end{figure*}

As a check on our automated measurements, we manually measured, the plane of sky speeds, length, width, and lifetime of 50 randomly selected network microflashes (Figure \ref{fig5}) from 29-March-2023. Their speeds fall  in the range of 10--40 \kms, with a mean speed of 28 $\pm$ 11.5 \kms, which is of the  order of the chromospheric sound speed. The length and width of most of the network microflashes are from 350 to 600 km and from 250 to 350 km, respectively.  The lifetime of the majority of manually measured microflashes  is in between 8 and 80 seconds, with a mean of 30 $\pm$ 15 seconds.

\subsection{Wavelet Analysis} \label{wavelet}

\begin{figure*}[h]
	\centering
	\includegraphics[width=\linewidth]{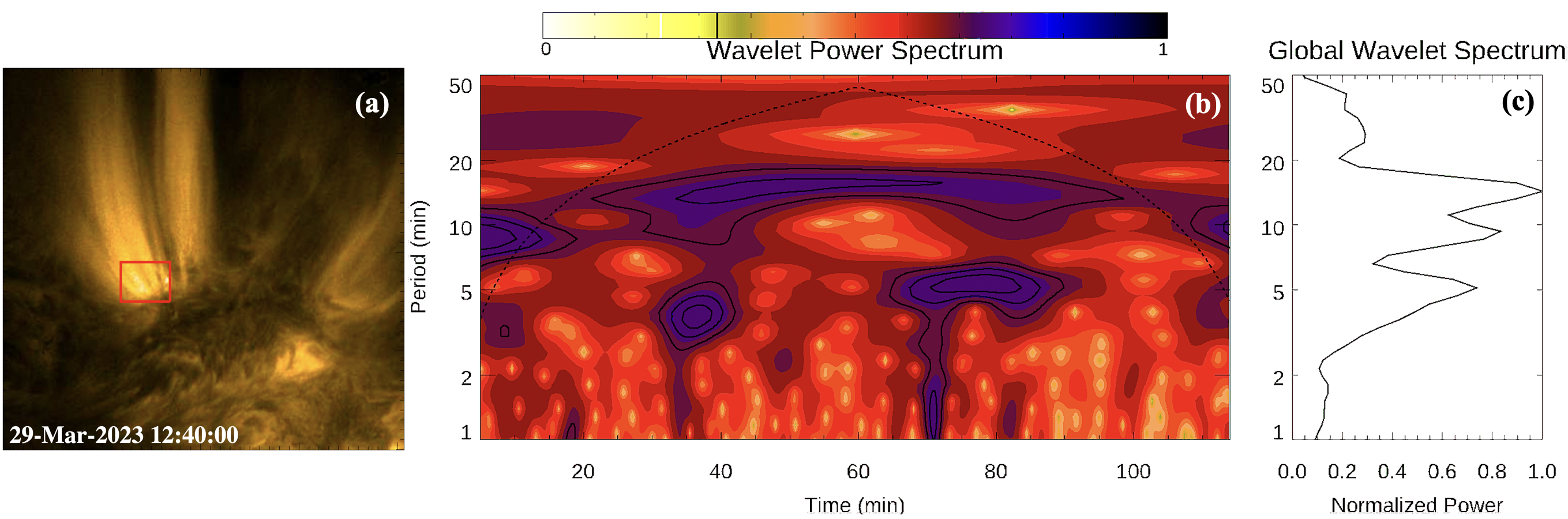} 
	\caption{Results from the wavelet analysis. Panel (a) shows an \hri\ 174 \AA\ image and the red box shows the region that was analyzed. Panel (b)  shows the normalized wavelet power spectrum, inside the red box of (a), as function of time and period. The wavelet spectrum shows the power as a function of time and period, over the time of the \hri\ observations. The black dashed curve depicts the cone of influence the values outside of which are subject to edge effects and are considered unreliable. Panel (c) plots the normalized power-spectrum given by time integral of  panel (b). 
	}  \label{A2}
\end{figure*}

To examine the periodicity of  microflashes, we performed  a Morlet wavelet analysis \citep{torrence1998} on \hri\ 174 \AA\ images.  We first obtain the maximum intensity at each time step from a box of size 50 by 40 pixels, drawn at the plume base (shown in the red box in Figure \ref{A2}a). We subtract the mean from this time series. Then, we apply a 15-minute box-car smoothing to the mean-subtracted time series.  To remove longer-period power from the mean-subtracted time series, we subtract from the mean-subtracted time series the smooth curve given by the 15-minute box-car smoothing.  Finally, to the remaining time series, we apply the wavelet power-spectrum analysis to obtain the power spectrum shown in Figure \ref{A2}. The wavelet analysis used Morlet wavelet kernels that ranged in full-width-half-maximum both above and below 10 minutes by roughly a factor of 5.

Figure \ref{A2}b shows the wavelet power spectra for 174 \AA\  brightenings in the red box in  Figure \ref{A2}a. The blue areas in  Figure \ref{A2}b show the periods and times of the highest power, which  peaks at $\sim$ 5, 10, and 15 minutes. 

\section{Data set 2}\label{ex-2}

\begin{figure*}[ht]
	\centering
	\includegraphics[width=\linewidth]{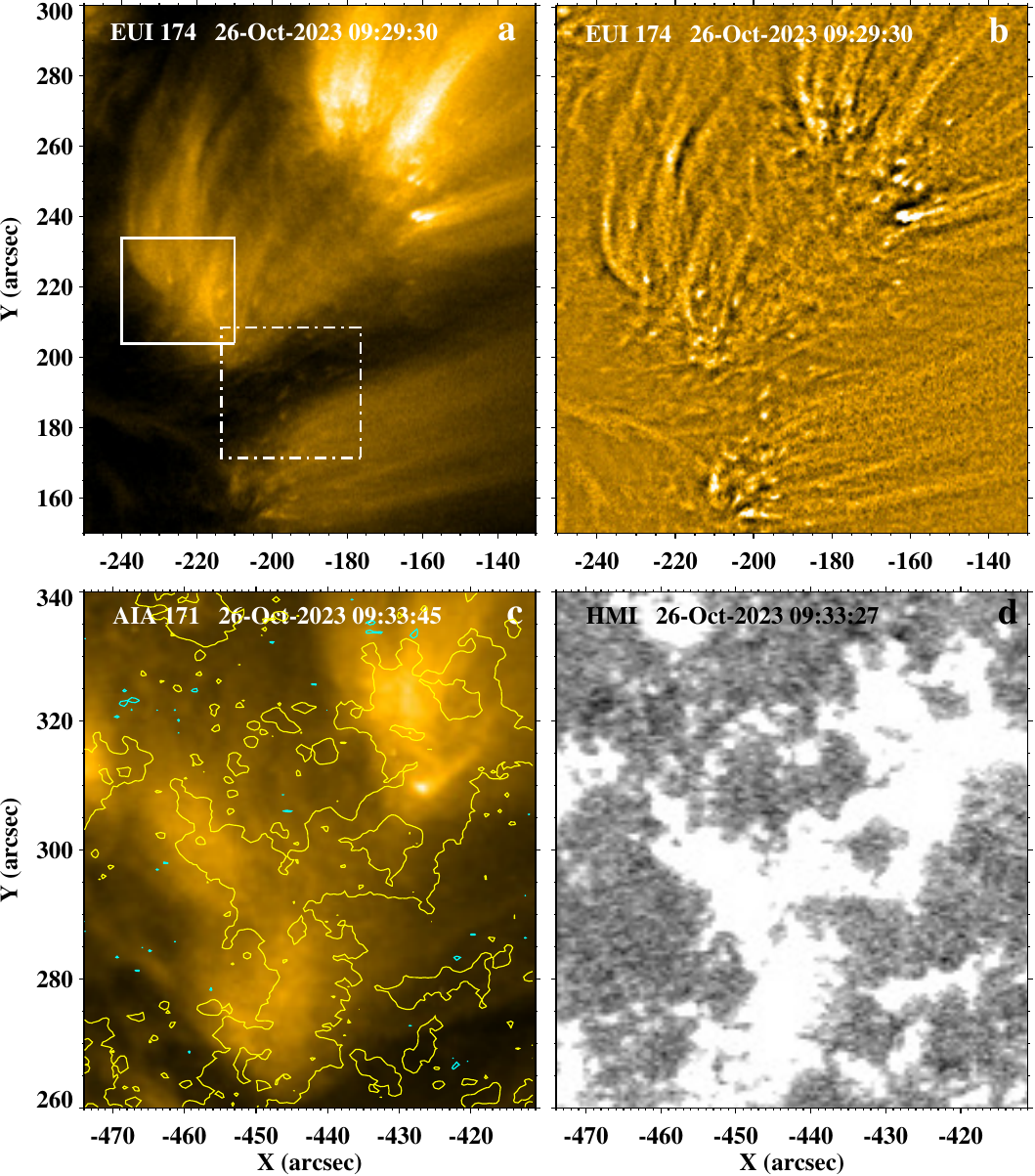} 
	\caption{An example of network microflashes at a plume-base observed by \hri\ and SDO on 26-October-2023. Panel (a) shows \hri\ 174 \AA\ image of solar plumes and microflashes. Panel (b) shows the unsharp mask version of the same image. Panel (c) shows the same FOV in AIA 171 \AA. Panel (d) shows the line-of-sight photospheric magnetic flux in the same FOV.  In panel (c), HMI contours, of levels $\pm$20 G, at 09:33:271 UT are overlaid, where yellow and cyan contours outline positive and negative magnetic flux, respectively.  The solid white box and dotted-dashed white box in (a) outlines the FOV in Figures \ref{fig2} and \ref{fig2a}, respectively. The animation (Movie1) runs from 09:00 to 11:00 UT. The animation is unannotated and the FOV is same as in Panel (a). 
	}  \label{fig1}
\end{figure*}

\begin{figure*}[h]
	\centering
	\includegraphics[width=0.98\linewidth]{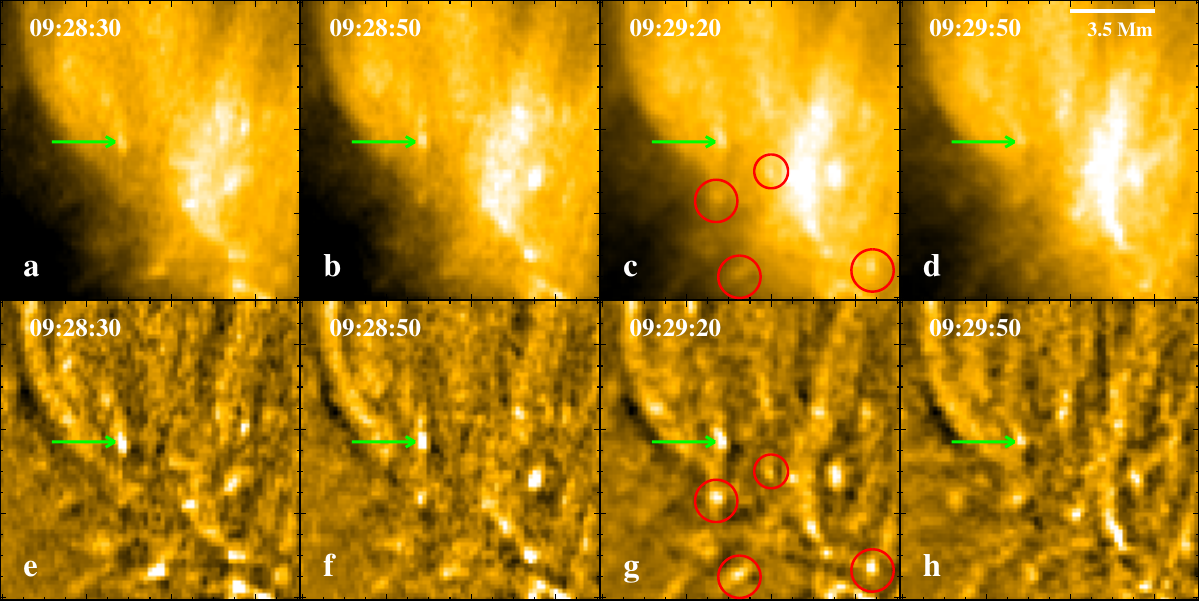}
	\caption{Network microflashes at the base of a plume. Panels (a--d) show 174 \AA\ \hri\ images of the plume base region in the FOV that is shown within the solid white box region of Figure \ref{fig1}a. Panels (e--g) show the unsharp mask images of the same. The green arrows point to a network microflash through most of its life. The red circles enclose other microflashes that are visible in these \hri\ frames.  
	}  \label{fig2}
\end{figure*}

\begin{figure*}
	\centering
	\includegraphics[width=\textwidth]{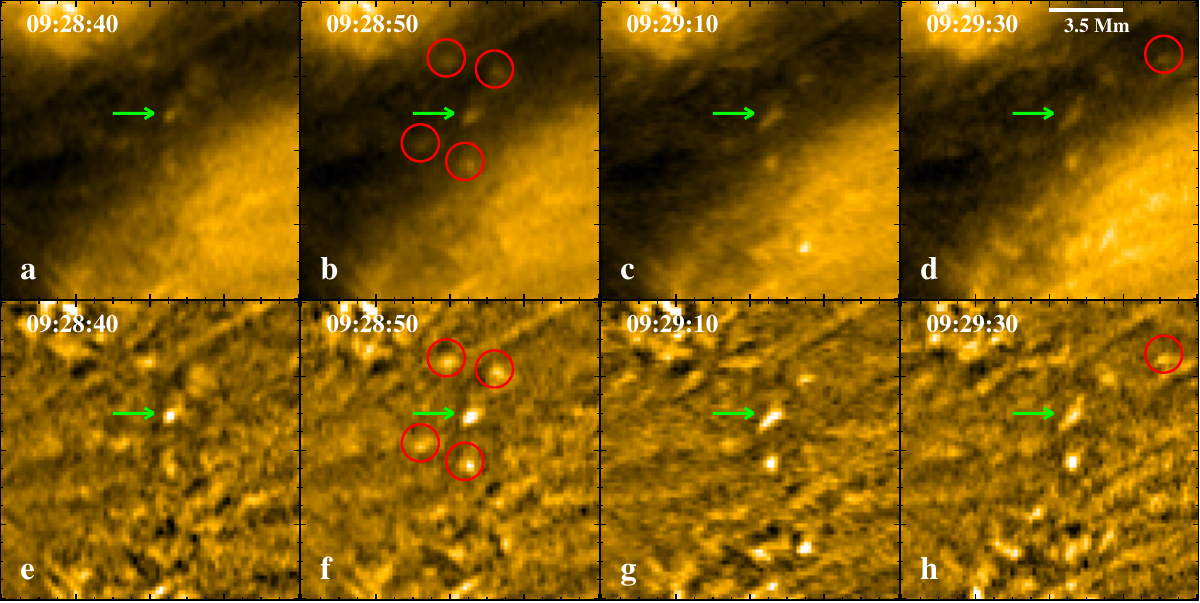}
	\caption{Network microflashes in network away from the plume base. Panels (a--d) show the 174 \AA\ \hri\ images in the FOV of the dotted-dashed white box region of Figure \ref{fig1}a. Panels (e--g) show the unsharp mask versions of those images. The green arrow points to a network microflash through most of its life. The red circles enclose other microflashes that are visible in these \hri\ frames.  
	}  \label{fig2a}
\end{figure*}

Figure \ref{fig1}a is an \hri\ image of plumes that shows many fine-scale brightenings at the base of the plume bushes. These bright features (network microflashes) are better visible in the unsharp masked image (Figure \ref{fig1}b and Movie2). The plumes are rooted in apparently unipolar (positive-polarity) magnetic flux clumps (Figure \ref{fig1}d). In Figures \ref{fig2} and \ref{fig2a}, we show zoomed in views of \hri\ microflashes at the base of a plume and in non-plume-base network flux, respectively.  The green arrow in Figure \ref{fig2} points to a microflash in the base of the plume. It follows the evolution of this microflash - from turn-on to fade-out.  The microflash is an elongated feature whose lifetime is about 80 seconds. Many microflashes are visible in these images. A few of them are enclosed in red circles. 

Similarly, Figure \ref{fig2a} shows  examples of  network microflashes in the non-plume-base positive network flux in the dotted-dashed white box region of Figure \ref{fig1}a. The HMI magnetogram in Figure \ref{fig1} displays the photospheric magnetic flux in the FOV of Figure \ref{fig1}. The FOV of Figure \ref{fig2a} centers on the lane of positive network flux that connects the two positive network flux clumps of two plume feet. That interval of network lane is not in the base of a plume. There are several network microflashes  in this non-plume-base interval of unipolar network-flux. The green arrows in Figure \ref{fig2a} follow most of the life of one of these microflashes. The microflash is an elongated bright feature that  extends along the direction of the plume field. The lifetime of this microflash is 50 seconds. This second example verifies that microflashes are present in unipolar network in and out of the base of plumes. Due to its slower cadence, we did not measure the microflashes from this dataset. 

\begin{figure}
	\includegraphics[width=0.4\linewidth]{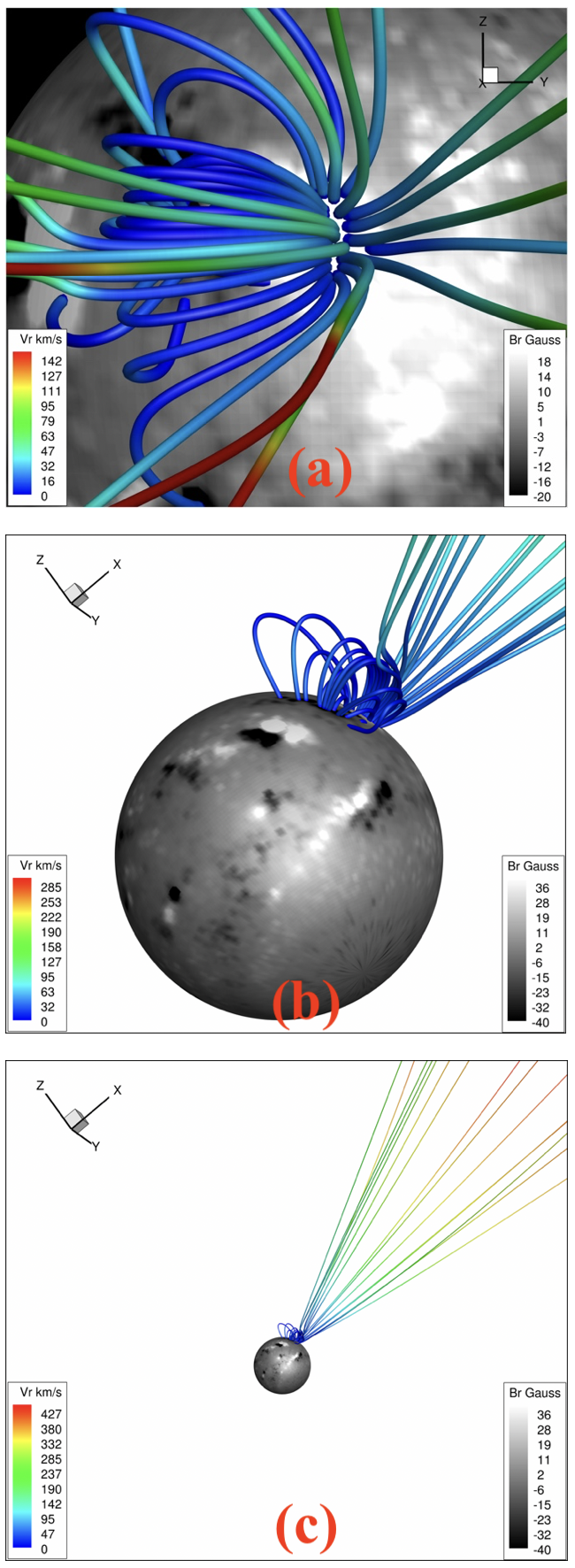}
	\caption{The MHD model 3D field from the plume region in the 2023 October 26 observations. The radial magnetic field is shown at r = 1.006 Rs with gray scale. Selected field lines are shown from the plume region. The color on the field lines shows the radial velocity of the solar wind. Panel (a) shows a similar point of view as the observation. Panels (b) and (c)  show a different point of view. The field lines in the second and third figures extend to $\sim$2 and $\sim$14 Rs, respectively. } \label{A3}
\end{figure}

Figure \ref{A3} shows the MHD model 3D magnetic field configuration (Methods) of the plume region of 26-Oct-2023. The model open magnetic field lines from this plume region are shown extending into the heliosphere. The MHD model gives fast solar wind coming from this plume region, which again might be driven by what makes network microflashes. 

%
%

\bibliographystyle{aasjournal}

\end{document}